\documentclass[doublecol,letterpaper,final]{epl2}
\usepackage{amssymb}
\usepackage{amsmath}
\usepackage{bbm}
\usepackage{graphicx}

\def\GroupeEquations#1{\begin{subequations}  #1  \end{subequations}}
\def\moy#1{\left\langle #1 \right\rangle}

\def\Tr{\text{Tr}}

\def\HF{Hirsch-Fye }  
 
\begin{document}

\author{Emanuel Gull \inst{1} \and Philipp Werner \inst{2} \and Olivier Parcollet \inst{3} \and Matthias Troyer \inst{1}}

\institute{                    
  \inst{1} Theoretische Physik, ETH Z{\"u}rich, 8093 Z{\"u}rich, Switzerland\\
  \inst{2} Columbia University, 538 West, 120th Street, New York, NY 10027, USA \\
  \inst{3} Institut de Physique Th{\'e}orique, CEA/DSM/IPhT-CNRS/URA 2306 CEA-Saclay, F-91191 Gif-sur-Yvette, France
}

\title{Continuous-time auxiliary field Monte Carlo for quantum impurity models}

\date{January 23, 2008}

\hyphenation{}

\abstract{
We present a continuous-time Monte Carlo method for quantum impurity models, which combines a weak-coupling expansion with an auxiliary-field decomposition. The method is considerably more efficient than Hirsch-Fye and free of time discretization errors, and is particularly useful as  
impurity solver in large cluster dynamical mean field theory (DMFT) calculations.   
}


\maketitle

\section{Introduction}

The development of efficient numerical methods for solving quantum impurity models has been driven in recent years 
by the success of dynamical mean field theory (DMFT)
\cite{Georges96,Georges92,Kotliar06} and its extensions.
DMFT is an approximate framework for the study of fermionic lattice models, 
which replaces the lattice by  a quantum impurity embedded in a self-consistent bath.
Both cluster-extensions of DMFT\cite{Hettler98, Maier06, Kotliar06,LichtensteinKatsnelson,Kotliar01}
and realistic electronic structure calculations, which combine DMFT with band structure methods \cite{Kotliar06},
involve multi-site or multi-orbital impurity models (e.g. for $d$- and $f$-electron systems),
whose solution is computationally expensive and in practice the bottleneck of the calculations. 
In order to facilitate progress in this field, it is therefore important to develop 
fast and accurate impurity solvers.

Until recently,  the Hirsch-Fye auxiliary field method\cite{Hirsch86} has been the only  Quantum Monte Carlo impurity solver used in DMFT. It suffers from one major drawback: it requires a discretization of the imaginary time interval into a large number $N$ of time slices and therefore the calculation of determinants of $Nn_s\times Nn_s$ matrices (for models with $n_s$ sites and on-site interactions only), which is computationally expensive. Furthermore, this discretization introduces a systematic error which needs to be dealt with (in principle) through tedious extrapolations $N\rightarrow \infty$. 

Important progress was achieved recently with the development of continuous-time impurity solvers, which are based on the stochastic sampling of a diagrammatic expansion of the partition function. These methods do not suffer from time discretization errors and allow the simulation of models with more general interactions. 
The first continuous-time impurity solver was proposed by Rubtsov {\it et al.} \cite{Rubtsov05}, who expanded the partition function in the interaction terms and used Wick's theorem.
Another powerful and flexible diagrammatic solver for small impurity problems, based on a diagrammatic expansion in the impurity-bath hybridization, has been proposed in 
Refs.~\cite{Werner05, Werner06, Haule07}.
Since this method perturbs around an exactly solved atomic limit, it is particularly efficient at moderate and strong interactions \cite{Gull07}.
While the sign problem in this algorithm is less severe than in Hirsch-Fye or in the weak-coupling continuous-time method, the computational effort scales exponentially with the number of sites and orbitals, making it difficult or impossible to solve clusters with eight or more sites. As a consequence, Hirsch-Fye is currently still considered the method of choice for large cluster DMFT computations. 

In this paper, we present a new continuous-time impurity solver
which combines a weak-coupling expansion with an auxiliary field
decomposition, and which was inspired by the work of Rombouts {\it et al.} \cite{Rombouts99} for lattice models.  Our method is formally similar to the Hirsch-Fye algorithm, but as a weak-coupling solver performs comparable to Rubtsov's method.

\section{Method}\label{sec:method}
\def\stau{\{ s_{i},\tau_{i}\}} 
We present the algorithm for the single-impurity model corresponding to the DMFT solution of the 
one-band Hubbard model. 
The extension to multi-band or cluster models with density-density
coupling is straightforward and will be briefly discussed at the end of this section. 

The partition function for the impurity model can be written as a path integral over Grassman variables $\xi$ and $\xi^*$, $Z=\int \mathcal{D}[\xi,\xi^*]e^{-S}$, with effective action
\begin{align}
S &= \int_{0}^{\beta} \! d\tau d\tau' \!\sum_{\sigma=\uparrow,\downarrow}\!
 \xi^*_\sigma(\tau)\Big[g_{0\sigma}^{-1}(\tau-\tau')\Big]\xi_\sigma(\tau')\nonumber\\
&+U\int_{0}^{\beta} d\tau \Bigl(
n_\uparrow(\tau)n_\downarrow(\tau)-\frac{n_\uparrow(\tau)+n_\downarrow(\tau)}{2}
\Bigr).
\label{S_g0}
\end{align}
Here, $n=\xi^*\xi$ and $g_{0\sigma}$ is related to the ``conventional" non-interacting Green's function of 
Ref.~\cite{Georges96}
 by the expression $g_0^{-1}(i\omega_n)=-(g_{0,\text{conv}}^{-1}(i\omega_n)-U/2)$, which means that the chemical potential is shifted by $-U/2$ and $g_0(\tau)>0$ for $0\le \tau\le\beta$. 

In order to closely follow the standard derivation of the Hirsch-Fye algorithm (see e.g. 
Ref.~\cite{Georges96}),
we switch to the Hamiltonian formulation 
\GroupeEquations{
\begin{eqnarray}
H &=& H_{0} + V, \\
H_0 &=& -(\mu-U/2)(n_\uparrow+n_\downarrow)\nonumber\\
&&+\sum_{\sigma, p} ({\tilde t}_{\sigma,p} c^\dagger_\sigma a_{p} + h. c.) + \sum_{\sigma,p} \epsilon_p a^\dagger_{p,\sigma}a_{p,\sigma},\\
V &=& U  (n_\uparrow n_\downarrow - (n_\uparrow +n_\downarrow)/2),
\end{eqnarray}
}
where $H_{0}$ is the Gaussian term containing both the impurity ($c$) and the bath ($a$) degrees of freedom.
Following Rombouts {\it et al.} \cite{Rombouts99} 
we introduce a constant $K$, express the partition function in an interaction representation,  
\begin{equation}
Z=\Tr e^{-\beta H} =e^{-K}\Tr \Big[e^{-\beta H_0}T_{\tau} e^{-\int_0^\beta d\tau ( V(\tau)-K/\beta)}\Big]
\label{Z1}
\end{equation}
and expand the time ordered exponential in powers of $K/\beta - V$ (dropping the irrelevant factor $e^{-K}$): 
\begin{align}
Z =& \sum_{n\geq 0}
 \int_0^\beta d\tau_1 \ldots  \int_{\tau_{n-1}}^\beta \!\!\!\! d\tau_n 
\Big( \frac{K}{\beta}\Big)^n  
\Tr \Big[ e^{ -(\beta-\tau_{n}) H_{0}} \nonumber\\
&\times\Big( 1- \frac{\beta V}{K}  \Big) 
\ldots
e^{ - (\tau_{2} -\tau_{1}) H_{0}} \Big( 1- \frac{\beta V}{K} \Big) 
e^{ -\tau_{1} H_{0}}
\Big]. 
\end{align}
We then decouple the interaction terms as follows \cite{Rombouts99}:
\GroupeEquations{
\label{decouple}
\begin{align}
1-\frac{\beta V}{K}&= \frac{1}{2}\sum_{s=-1,1}e^{\gamma s (n_\uparrow-n_\downarrow)},\label{expansion_parameter}\\
\cosh(\gamma)&\equiv 1+(\beta U)/(2K).\label{gamma}
\end{align}
}
Expressions (\ref{expansion_parameter}) and (\ref{gamma}) are valid for arbitrary (complex) parameters $K$. If $K>0$, $\gamma$ is real and the expansion parameter is positive. 
After the decoupling, the partition function is of the form
\GroupeEquations{
\label{Z2}
\begin{gather}
Z= \sum_{n\geq 0}
\sum_{s_i= \pm 1 \atop 1\leq i\leq n}
\int_0^\beta d\tau_1 \ldots  \int_{\tau_{n-1}}^\beta \!\!\!\! d\tau_n 
\Big( \frac{K}{2\beta}\Big)^n    Z_{n}(\stau), \\
Z_{n}(\stau) \equiv \Tr 
\prod_{i=n}^{1} 
\exp( - \Delta \tau_{i} H_{0} ) 
\exp (s_{i} \gamma (n_{\uparrow} - n_{\downarrow})  ),
\end{gather} }
%
with $\Delta \tau_{i} \equiv \tau_{i+1} - \tau_{i}$ for $i<n$ and $\Delta \tau_{n} \equiv \beta-\tau_{n}+\tau_{1}$.
$Z_{n}$ is very similar to the expression for the partition function in the Hirsch-Fye algorithm after the Trotter approximation, see for example
Eq.~(117) of 
Ref.~\cite{Georges96},
 except that the time arguments of the auxiliary spins $s_{i}$ are not regularly spaced on $[0,\beta]$.
Indeed, one can straightforwardly generalize the calculation in 
Ref.~\cite{Georges96}
 to rewrite $Z_{n}/Z_{0}$  ($Z_0=\Tr e^{-\beta H_0}$) as: 
\begin{eqnarray}
\frac{Z_{n}(\stau) }{Z_0} &=& \prod_{\sigma=\uparrow,\downarrow} \det N_\sigma^{-1}(\stau),
\label{Zn_div_Z0}\\
N^{-1}_{\sigma}(\stau) &\equiv& e^{V_{\sigma}^{\{s_{i}\}}} -G_{0\sigma}^{ \{\tau_{i}\}} 
\Big(e^{V_{\sigma}^{ \{s_{i}\} }} - 1 \Big),
\\
e^{V_{\sigma}^{\{s_{i}\}}}&\equiv&\text{diag}\Big(e^{\gamma (-1)^{\sigma }s_1}, \ldots, e^{\gamma (-1)^{\sigma } s_n}\Big),
\end{eqnarray}
with the notations $(-1)^{\uparrow}  \equiv 1$, $(-1)^{\downarrow}  \equiv -1$ and 
$(G_{0\sigma}^{ \{\tau_{i}\}})_{i,j}=g_{0\sigma}(\tau_i-\tau_j)$ for $i\neq j$, $(G_{0\sigma}^{ \{\tau_{i}\}} )_{i,i}=g_{0\sigma}(0^+)$.

\def\sumSTau {\sum_{n\geq 0} \Big(\frac{K}{2\beta}\Big)^n \sum_{s_i= \pm 1 \atop 1\leq i\leq n}
\int_0^\beta d\tau_1 \ldots  \int_{\tau_{n-1}}^\beta \!\!\!\! d\tau_n }

While we tried to emphasize in our derivation the similarities to the Hirsch-Fye algorithm, let us note at this point also the essential differences between Hirsch-Fye and our continuous-time auxiliary field method (CT-AUX): 
{\it i)} CT-AUX is based on a weak-coupling expansion, not a Suzuki-Trotter decomposition of the partition function;
 {\it ii)} the auxiliary fields in CT-AUX originate from Rombout's decoupling formula (\ref{expansion_parameter}).  
In particular, CT-AUX does not require any time discretization. The number and position of auxiliary spins on the imaginary time interval is arbitrary and changes constantly during the simulation.

The formulae above are easily generalized for cluster and
multiorbital DMFT problems with density-density interactions  
by performing a similar expansion for all the interaction terms. 
For clusters of size $n_s$ with local density-density interaction $U$ (relevant e.g. for cluster DMFT 
approximations of the Hubbard model), 
expression (\ref{gamma}) for $\gamma$ remains unchanged and
other formulas, 
like Eq.~(\ref{Zn_div_Z0}), 
can be generalized straightforwardly by replacing the 
time and spin indices by (time, site) and (spin, site) multi-indices respectively.
The multiplicative factor dropped from the partition function is $\exp(-K n_s)$ in this case.

\subsection{Sampling procedure}

Our algorithm samples time ordered configurations consisting of spins $s_1, \ldots, s_n$ at times $\tau_1 <\tau_2<\ldots <\tau_n$ with weight
\begin{equation}
w(\stau)=\Big(\frac{K d\tau}{2\beta}\Big)^n \prod_{\sigma=\uparrow,\downarrow} \det N_\sigma^{-1}(\stau).
\end{equation}
For ergodicity it is sufficient to insert/remove spins with random orientation at random times. 

The detailed balance condition 
 can be implemented as follows. Assuming that we pick a random time in the interval $[0,\beta)$ and a random direction for this new spin ($p^\text{prop}(n\rightarrow n+1)=(1/2)(d\tau/\beta$)), and propose to remove it with probability $p^\text{prop}(n+1\rightarrow n)=1/(n+1)$, we get
\begin{equation}
\frac{p(n\rightarrow n+1)}{p(n+1\rightarrow n)}=\frac{K}{n+1}\prod_{\sigma=\uparrow,\downarrow} \frac{\det (N^{(n+1)}_\sigma)^{-1}}{\det (N^{(n)}_\sigma)^{-1}}.
\end{equation}

The matrices $N_\sigma=(e^{V_\sigma}-G_{0\sigma}(e^{V_\sigma}-I))^{-1}$ are stored and manipulated using fast update formulas analogous to those of 
Refs.~\cite{Rubtsov05, Werner05}.
 When inserting a spin we add a new row and column to $N_\sigma^{-1}$. Following the notation of 
 Ref.~\cite{numrec92},
  we define the blocks  (omitting the $\sigma$ index until the end of this section) 
\begin{align}(N^{(n+1)})^{-1} = \begin{pmatrix}(N^{(n)})^{-1}&Q\\R&S\end{pmatrix} ,\hspace{1mm} N^{(n+1)}=\begin{pmatrix}\tilde P&\tilde Q\\\tilde R&\tilde S\end{pmatrix},\label{Nmatrix}\end{align}
where $Q$, $R$, $S$ denote $(1\times n)$, $(n\times 1)$, $(1\times1)$ matrices, respectively, which contain the contribution of the added spin. The determinant ratio needed for the acceptance/rejection probability is then given by
\begin{align}\frac{\det (N^{(n+1)})^{-1}}{{\det (N^{(n)}})^{-1}} = \frac{1}{\det \tilde S} = S-R (N^{(n)})Q.\end{align}
As we store $N^{(n)},$ computing the acceptance/rejection probability of an insertion move involves a matrix-vector multiplication followed by an inner product, i.e. an $O(n^2)$ operation.
If a move is accepted, a rank one update is performed to compute the new matrix $N^{(n+1)}$ out of  $N^{(n)}, Q, R$, and $S$:
\GroupeEquations{
\begin{align} 
 \tilde S &= (S-R [N^{(n)}Q])^{-1}, \\
 \tilde Q &= -[N^{(n)}Q] \tilde S, \\
 \tilde R &= -\tilde S [R N^{(n)}], \\
 \tilde P &= N^{(n)} + [N^{(n)}Q]\tilde S [R N^{(n)}].
\end{align}}

\subsection{Measurement of the Green's function}

The main observable of interest in the simulations is the Green's function $g_\sigma(\tau,\tau')$.
First, let us note from (\ref{Z2}) that one can add two additional ``non-interacting'' 
spins $s=s'=0$ at any fixed times $\tau$ and $\tau'$ (we denote with a tilde the corresponding matrices of size $n+2$).
$Zg_{\sigma}(\tau,\tau')$ is then given by an expression similar to
Eqs.~(\ref{Z2}), with an insertion of $c(\tau)$ and $c^{\dagger}(\tau')$ at
the corresponding times.
We can again use the standard Hirsch-Fye formula for the discretized
Green function (Eq.~(118) of 
Ref.~\cite{Georges96}) to obtain 
\begin{align}
\label{Gfinal}
g_{\sigma} (\tau,\tau')
=&
\frac{1}{Z}
\displaystyle \sumSTau \nonumber\\
&\times Z_n (\stau)  
\tilde G_{\sigma} ^{\stau}(\tau,\tau'),
\end{align}
with 
$\tilde G_{\sigma} ^{\stau} = 
\tilde N_{\sigma}(\stau) \tilde G_{0\sigma}^{ \{\tau_{i}\}}.$
%
Since $s=s'=0$, 
a simple block calculation leads to  
\begin{align}\label{GSreduc}
&\tilde G_{\sigma} ^{\stau}(\tau,\tau') =  g_{0\sigma} (\tau, \tau')\nonumber\\
&\hspace{4mm}+\sum_{k,l=1}^{n}
 g_{0\sigma} (\tau,\tau_{k})
\Big[
 (e^{V_{\sigma}^{\{s_{i}\}} } -1)N_{\sigma}(\stau)
 \Big]_{kl}
 g_{0\sigma}(\tau_l,\tau').
\end{align}

In order to compute the Green's function, 
one can not just accumulate its values at the discrete times $\tau_{i}$
of the auxiliary spins, since the $\{\tau_{i}\}$ are correlated.
Rather, the Green's function is accumulated using Eqs.~(\ref{Gfinal}) and (\ref{GSreduc}):
\begin{gather}
g_\sigma(\tau) = g_{0\sigma}(\tau)+ 
\int_{0}^{\beta} d \tilde \tau  g_{0\sigma}(\tau - \tilde \tau) 
\Big\langle S^{\stau}_{\sigma} (\tilde \tau)\Big\rangle,
\label{g_reduc}
\\ 
S^{\stau}_{\sigma} (\tilde \tau)  \equiv \sum_{k=1}^{n} 
\delta(\tilde \tau - \tau_{k})   \sum_{l=1}^{n} M_{kl}^{\stau} g_{0\sigma} (\tau_l),
\\
M_{kl}^{\stau} \equiv \big[(e^{V^{\{s_i\}}_\sigma}-1)N_\sigma(\stau)\big]_{kl},
\label{M_ij}
\end{gather}
where we have used translational invariance, set $\tau'=0$, and denoted the Monte Carlo 
average with angular brackets (our convention is $g(\tau)>0$ for $0<\tau<\beta$). 
Hence, we measure only the quantity  $\langle S^{\stau}_{\sigma} (\tilde \tau)\rangle$, which we bin into fine bins. 
After the simulation is completed, the Green's function is constructed using Eq.~(\ref{g_reduc}).

Note that the Dyson equation
\begin{equation}
\label{defDyson}
g_{\sigma}(i \omega_{n})  = g_{0\sigma}(i \omega_{n})  + g_{0\sigma}(i \omega_{n}) \Sigma_{\sigma}(i \omega_{n}) g_{\sigma}(i \omega_{n}) 
\end{equation}
implies that this procedure amounts to accumulating $\Sigma_{\sigma} g_{\sigma}$.
Besides the higher efficiency with respect to the direct accumulation of the Green's function,
 an important advantage of such a measurement is the reduction in high-frequency noise by the multiplication with $g_0\sim
1/\omega_n$ 
(see also Ref.~\cite{Bulla}
 for similar ideas in the
NRG-DMFT context).

Let us emphasize that the same procedure can also be employed in the weak
coupling algorithm, where it yields significant performance gains over
the methods described in 
Refs.~\cite{Rubtsov05} and
\cite{Gull07}, especially for large clusters.

\subsection{Four point functions}
 
Four point correlation functions 
can also be computed in a similar way as in \HF using the fact that for a fixed auxiliary spin configuration
the problem is Gaussian and Wick's theorem can therefore be used together with Eq.~(\ref{GSreduc}).
Thus the problem reduces to the accumulation of  
the determinant of a $2\times2$ matrix
\begin{equation}
\left\langle 
\left | 
\begin{matrix} 
(g_0^{12}+g_0^{1k}M_{kl}^{\stau}g_0^{l2}) & (g_0^{14}+g_0^{1k}M_{kl}^{\stau}g_0^{l4}) \\
(g_0^{32}+g_0^{3k}M_{kl}^{\stau}g_0^{l2}) & (g_0^{34}+g_0^{3k}M_{kl}^{\stau}g_0^{l4})
\end{matrix}
\right | 
\label{fpdet}\right\rangle
\end{equation}
with $M_{kl}^{\stau}$ defined in Eq.~(\ref{M_ij}).
If only a few correlation functions are measured, Eq.~(\ref{fpdet}) is best evaluated directly during the simulation. If many or all correlation functions have to be measured at $n_\tau$ time points and the size $n_M$ of $M$ is comparatively small, it is advantageous to accumulate only
 $\langle M_{ij}^{\stau}\rangle$ and $\langle M_{ij}^{\stau}M_{kl}^{\stau}\rangle$ 
 and reconstruct the correlation function at the end of the computation.
Indeed, while binning the latter expression is $O(n_\tau^3)$ in memory, it is only $O(n_{M}^3)$ computationally (using time translation invariance).


\begin{figure}[t]
\begin{center}
\includegraphics[width=0.9\columnwidth]{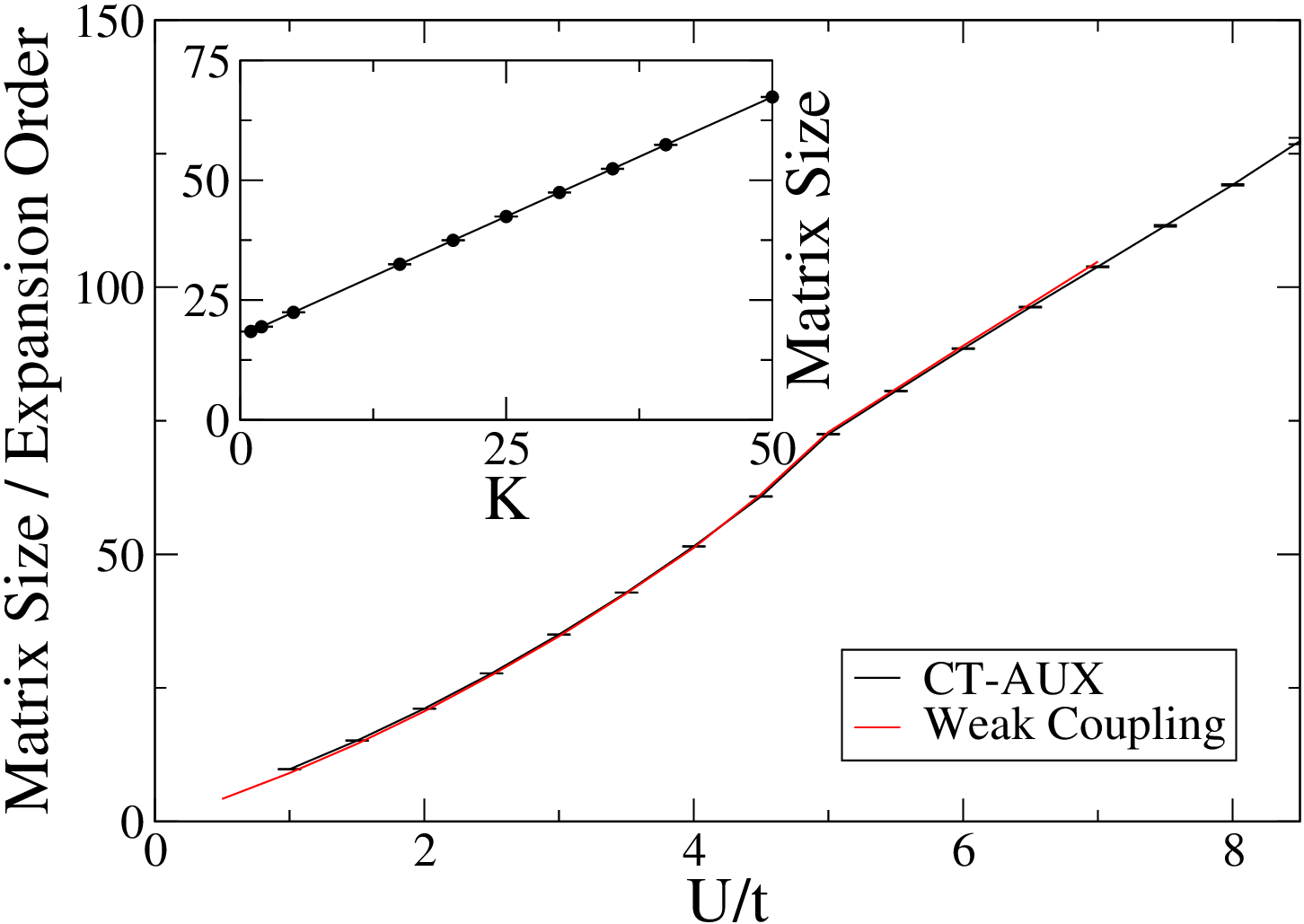}
\caption{Average perturbation order for the continuous-time auxiliary field algorithm ($K=1$) and the weak-coupling algorithm (with $\alpha=0.01$). Single-site Hubbard model, half-filling, semi-circular density of states of bandwidth $4t$, and $\beta t=30$. 
Inset: Expansion order (matrix size) as a function of $K$. Single site Hubbard model, half-filling, semicircular density of states of bandwidth $4t$,  $U/t=4$, and $\beta t=10$.
}
\label{order_combined}
\end{center}
\end{figure}



\subsection{Role of the expansion parameter $K$}
 
The average perturbation order $\langle n_{\text{ctaux}} \rangle$ is related to the parameter $K$, potential energy and filling by
\begin{equation}
\langle n_{\text{ctaux}} \rangle = K-\beta U \langle n_\uparrow n_\downarrow -(n_\uparrow+n_\downarrow)/2 \rangle. 
\label{order}
\end{equation}
This expression is obtained by applying the operator $\left. K\partial_{K} \right |_{U/K}$ 
to $\ln Z$ both in its original form (\ref{Z1}) and to (\ref{Z2}), including the factor $e^{-K}$ dropped after
Eq.~(\ref{Z1}) (see also 
Ref.~\cite{Rombouts99}).
In the case of the weak-coupling algorithm \cite{Rubtsov05}, $\langle n_\text{wc} \rangle_{\alpha\rightarrow 0} = -\beta U \langle n_\uparrow n_\downarrow -(n_\uparrow+n_\downarrow)/2 \rangle$, 
where $\alpha$ is the small parameter which must be introduced to reduce the sign problem. 
Hence, the perturbation order in the continuous-time auxiliary field method grows linearly with $K$ (see inset of Fig.~\ref{order_combined}) and $\langle n_{\text{ctaux}} \rangle_{K\rightarrow 0} = \langle n_\text{wc} \rangle_{\alpha\rightarrow 0}$. 

Figure~\ref{order_combined} shows the perturbation orders for the two methods as a function of $U$. For these small values of $K$ and $\alpha$, the perturbation orders are essentially identical. Both weak-coupling methods scale roughly linearly with $U$, with a kink visible at the Mott critical value. It also follows from Eq.~(\ref{order}) that the perturbation order is essentially linear in the inverse temperature $\beta$. 

Similar to the weak-coupling expansion parameter $\alpha$ \cite{Rubtsov05}, the parameter $K$ can be freely adjusted. 
While a larger $K$ yields a larger expansion order, it also reduces the value of $\gamma$ (see Eq.~(\ref{decouple})). This makes it easier to flip auxiliary spins. Therefore 
the auxiliary spins have less tendency to polarize for larger $K$. In practice, however, $K$-values of order $1$ turned out to be adequate. Although we found that the sign problem improves slightly with larger $K$, this small gain is more than compensated by the increase in computational cost at larger values of $K$.

\begin{figure}[t]
\begin{center}
\includegraphics[width=0.9\columnwidth]{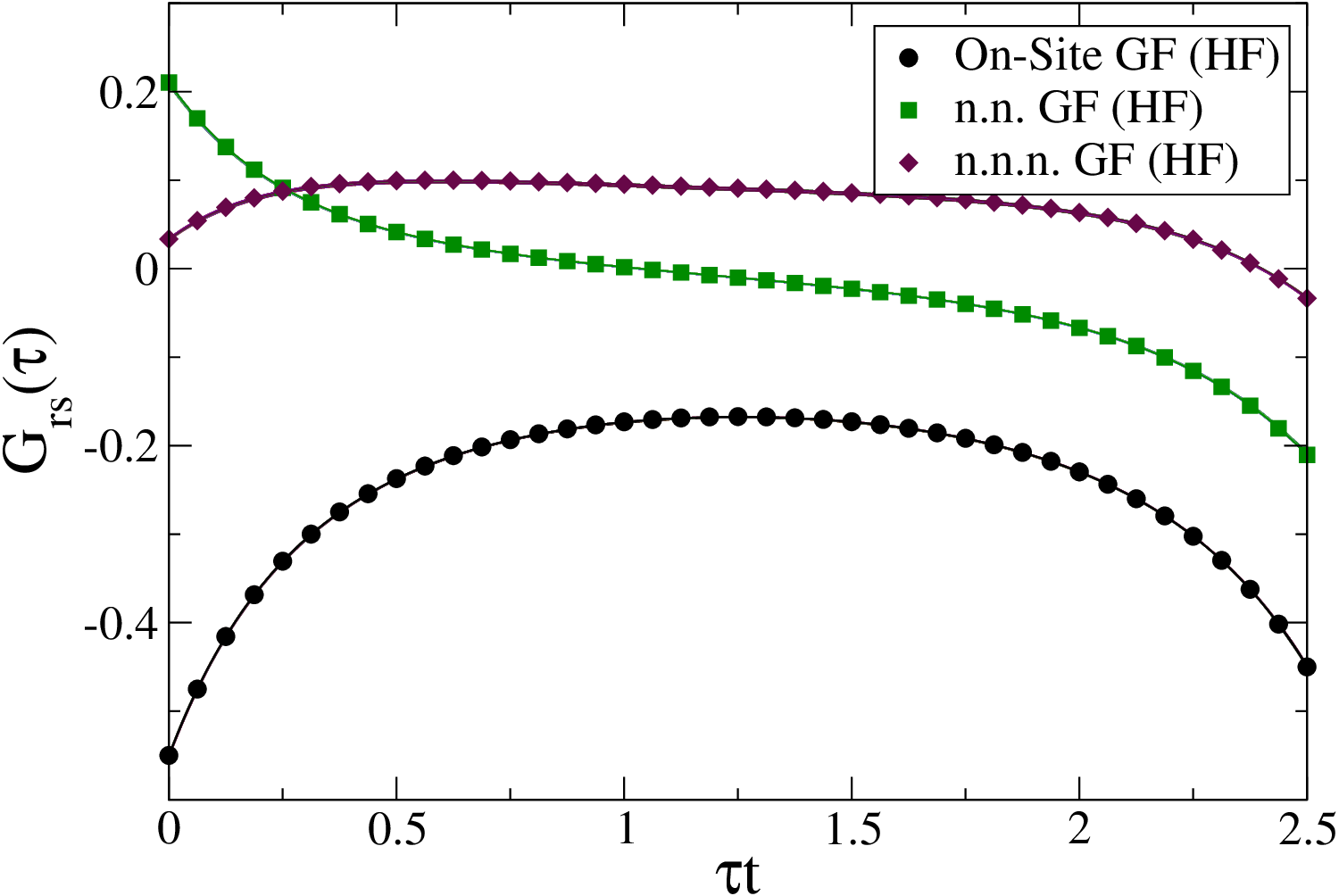}\\ 
\vspace{5mm}
\includegraphics[width=0.9\columnwidth]{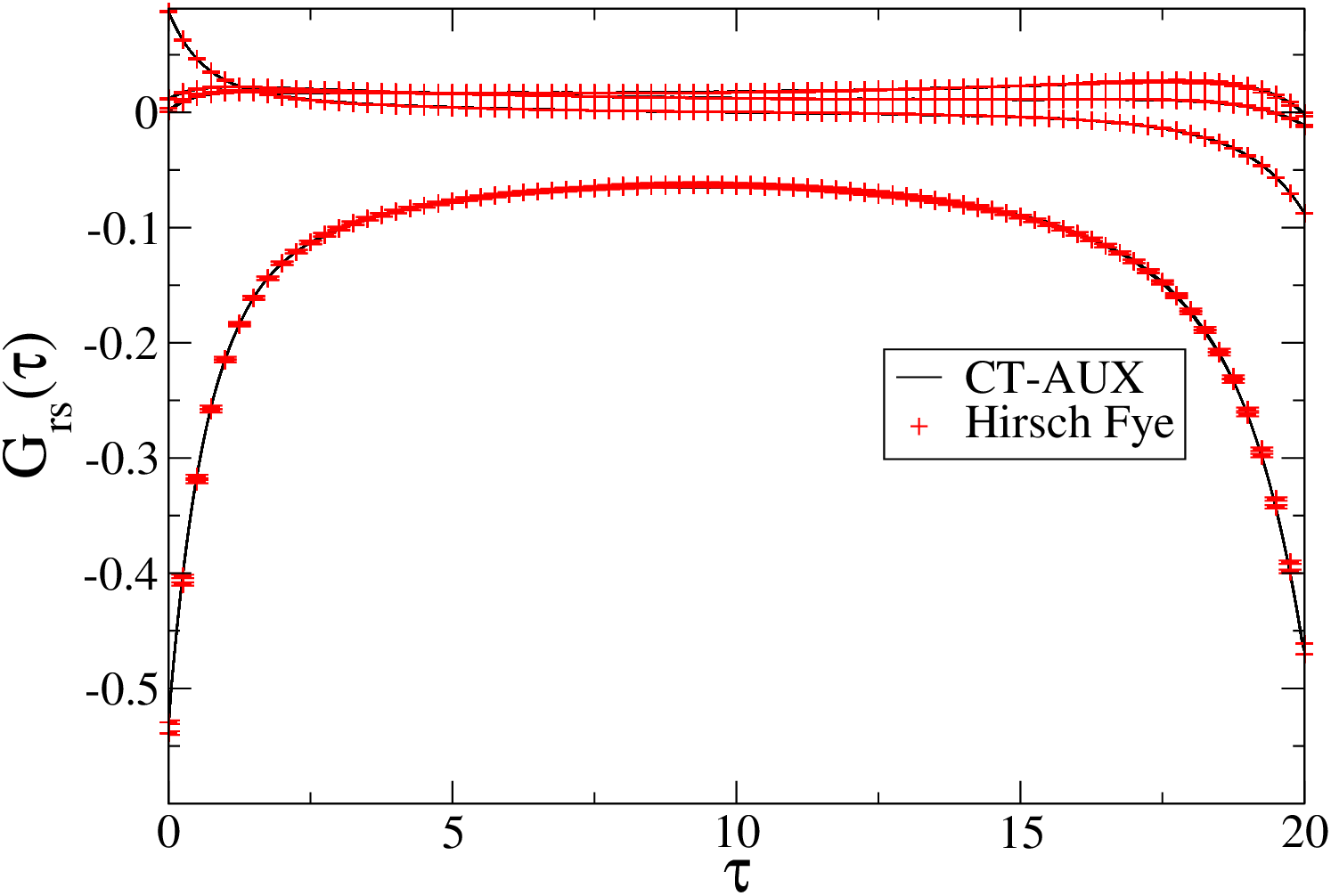}
\caption{Upper panel: real-space Green's functions (onsite, nearest-neighbor and next-nearest neighbor) for the four site cluster with nearest neighbor hopping $t$, $U/t = 4,$ filling = $0.9$, $\beta t=2.5$. Hirsch Fye results with 40 time slices are represented by the symbols, the weak coupling and CT-AUX results by lines (on top of each other).
Lower panel: real space Green's functions for the eight-site cluster obtained in 8 CPU hours using CT-AUX and Hirsch-Fye. 80 time slices were considered in the Hirsch-Fye simulation. Note the fact that the Hirsch-Fye result is slightly spin-polarized.}
\label{comparison_cluster_gf}
\label{GF_8site}
\end{center}
\end{figure}

\section {Comparison with other QMC methods}\label{sec:results}
To compare to other methods we implemented single-site, as well as 4 and 8 site cluster calculations in the dynamical cluster approximation (DCA)\cite{Hettler98, Hettler00}, and expect similar results for other cluster schemes such as cellular dynamical mean field theory 
(CDMFT) \cite{Kotliar01}.
The upper panel of Fig.~\ref{comparison_cluster_gf} shows a typical real space cluster Green's function for a 4-site DCA calculation. The CT-AUX results are identical to 
the other QMC results, showing the accuracy of the new approach.  The lower panel of Fig.~\ref{GF_8site} shows Green's functions for an 8-site cluster ($\beta =20, t=0.25, U=2, \mu = -0.3757$) obtained from a converged $g_0$ in 8 CPU hours on a 1.6 GHz Opteron 244. Symbols show the result for Hirsch-Fye with 80 time slices, and the lines indicate the result from CT-AUX, measured at 500 time points.

As a continuous time method CT-AUX has a definite advantage over 
Hirsch-Fye QMC, since it removes the necessity of careful extrapolations to the continuous time limit.
However, in order to be a useful replacement for Hirsch-Fye in practice, 
CT-AUX has to satisfy two requirements:
{\it i) }  The average expansion order
 $\moy{n}$, which determines the complexity of the calculation 
($O(\moy{n}^{3})$), has to be smaller than the number of times slices required  
in Hirsch-Fye (close to the continuous limit, where extrapolation is meaningful);
{\it ii) }  The sign problem must not be worse than in previous algorithms.

\subsection{Expansion order}

\begin{figure}[t]
\begin{center}
\includegraphics[width=0.9\columnwidth]{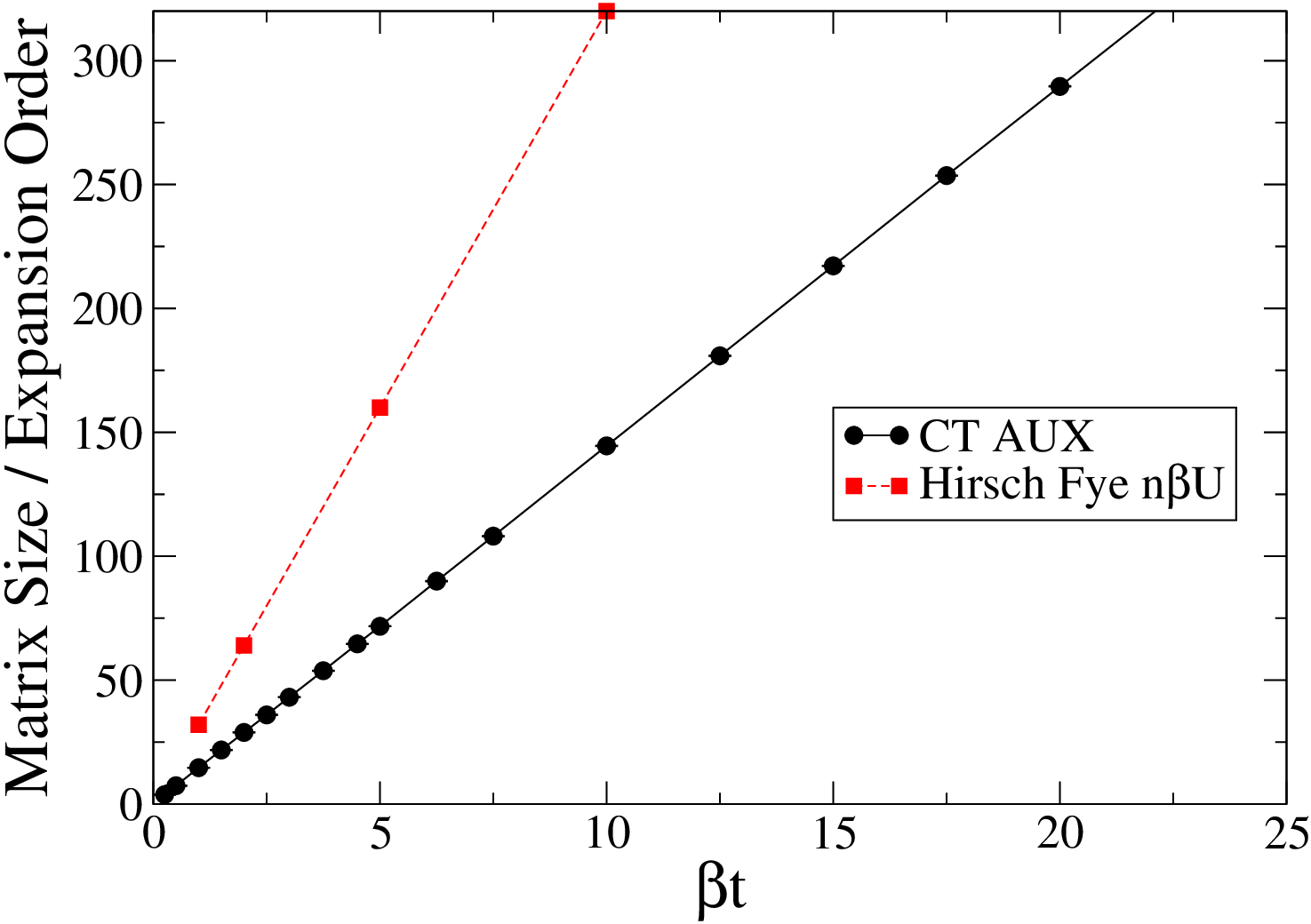}
\caption{Expansion order as a function of $\beta$ for the four site cluster with nearest neighbor hopping, $U = 2, \mu = -0.3757, t=0.25$. For Hirsch-Fye, a reasonable compromise between accuracy and speed would require at least $N=\beta U n_s$ time slices, which leads to larger matrices whose determinants need to be updated.}
\label{U2_t0.25_order_beta}
\end{center}
\end{figure}
First we compare the expansion order to Hirsch-Fye for a high temperature 
2x2 DCA calculation in Fig.~\ref{U2_t0.25_order_beta}.
In \HF  the number of time slices was fixed a priori using the optimistic criterion $N=\beta U n_s$ which corresponds to $\Delta {\tau} U =1$, where $\Delta \tau$ is the size of the time slices -- just barely in the region of validity of the Trotter approximation underlying the Hirsch-Fye method. CT-AUX with its roughly two times lower average perturbation order is much more efficient since both algorithms scale like the cube of the matrix size. 
In the 8-site cluster simulation of Fig.~\ref{GF_8site}, the \HF algorithm with 80 time slices had to update matrices of size 640, while CT-AUX merely had to operate on matrices of average size 136. This means that CT-AUX allows to reach substantially lower temperatures, even with modest computational resources. 

\begin{figure}[t]
\begin{center}
\includegraphics[width=0.9\columnwidth]{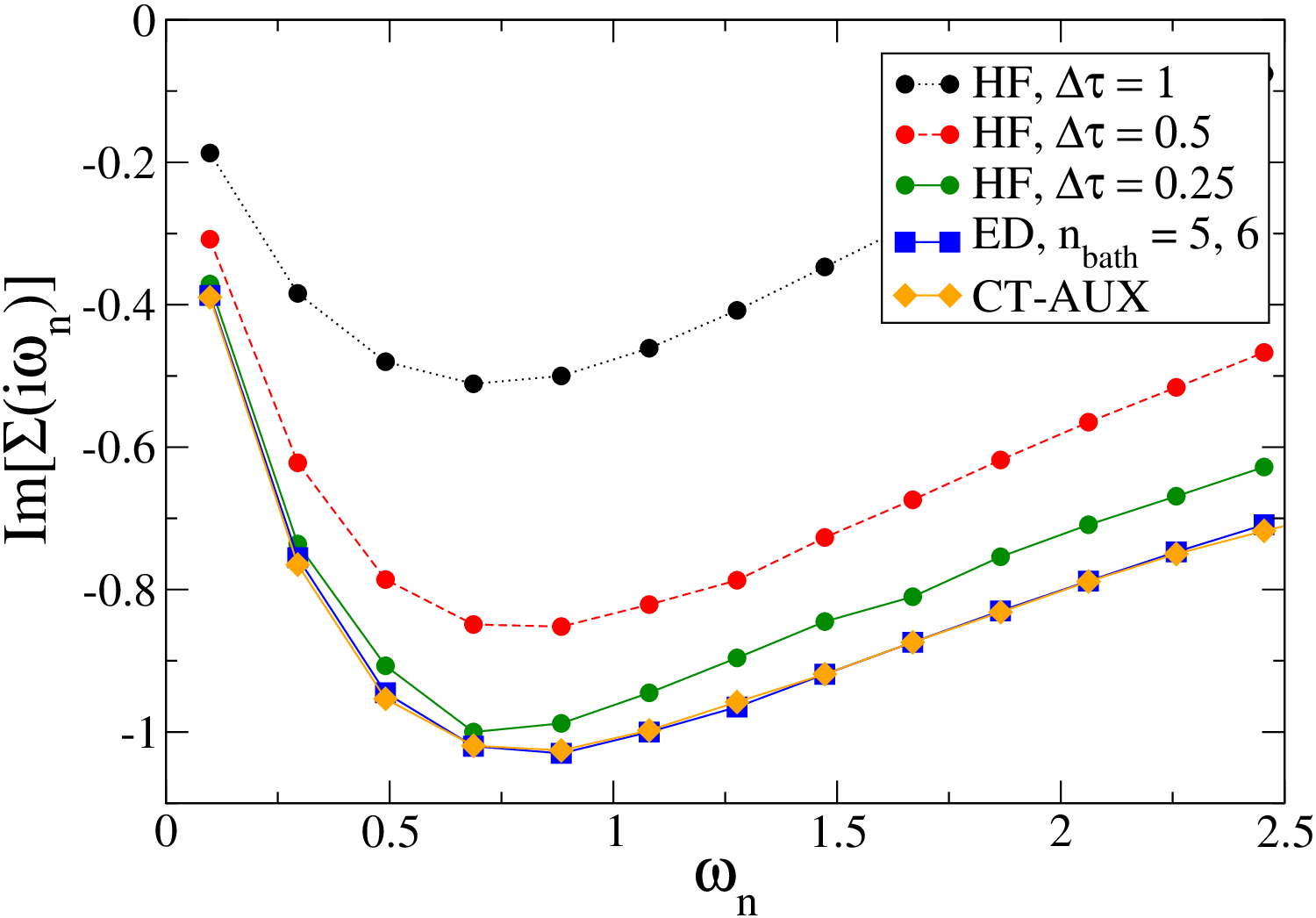}
\caption{
Imaginary part of the self-energy for the DMFT solution of the single site Hubbard
model. CT-AUX, Hirsch-Fye using 32, 64 and 128 auxiliary spins (time slices), ED with 6 bath sites ($\beta = 32$, $U=3$).
Hirsch-Fye and ED results were taken from
Fig.~15 of  Ref.~\cite{Georges96}.
For CT-AUX, the average number of auxiliary spins is $\moy{n} = 42.5$.}
\label{fig:Fig15bis}
\end{center}
\end{figure}

To make the comparison with \HF more precise and get rid of the
arbitrariness of the choice of the number of times slices in \HF 
we have reproduced in Fig.~\ref{fig:Fig15bis}
the self-energy calculation presented in Fig.~15 of 
Ref.~\cite{Georges96},
where the same single-site calculation was performed with \HF (32, 64 and 128 
time slices) and with exact diagonalization (ED). Even with 128 times slices, the \HF results still have 
substantial discretization errors while  CT-AUX produces the exact result (comparable to the $n_\text{bath}=6$ ED result) with only $\moy{n} = 42.5$ spins. 
This shows that CT-AUX is indeed much more efficient than the \HF method: not only does it compute the numerically exact result directly, but it does so using significantly less auxiliary spins.
This is due to the fact that CT-AUX, like Rubtsov's method, is based on a weak-coupling expansion (see Fig.~\ref{order_combined} and 
Ref.~\cite{Gull07}
 for a comparison of the weak-coupling method with Hirsch-Fye). 

\subsection{Sign problem}

\begin{figure}[t]
\begin{center}
\includegraphics[width=0.9\columnwidth]{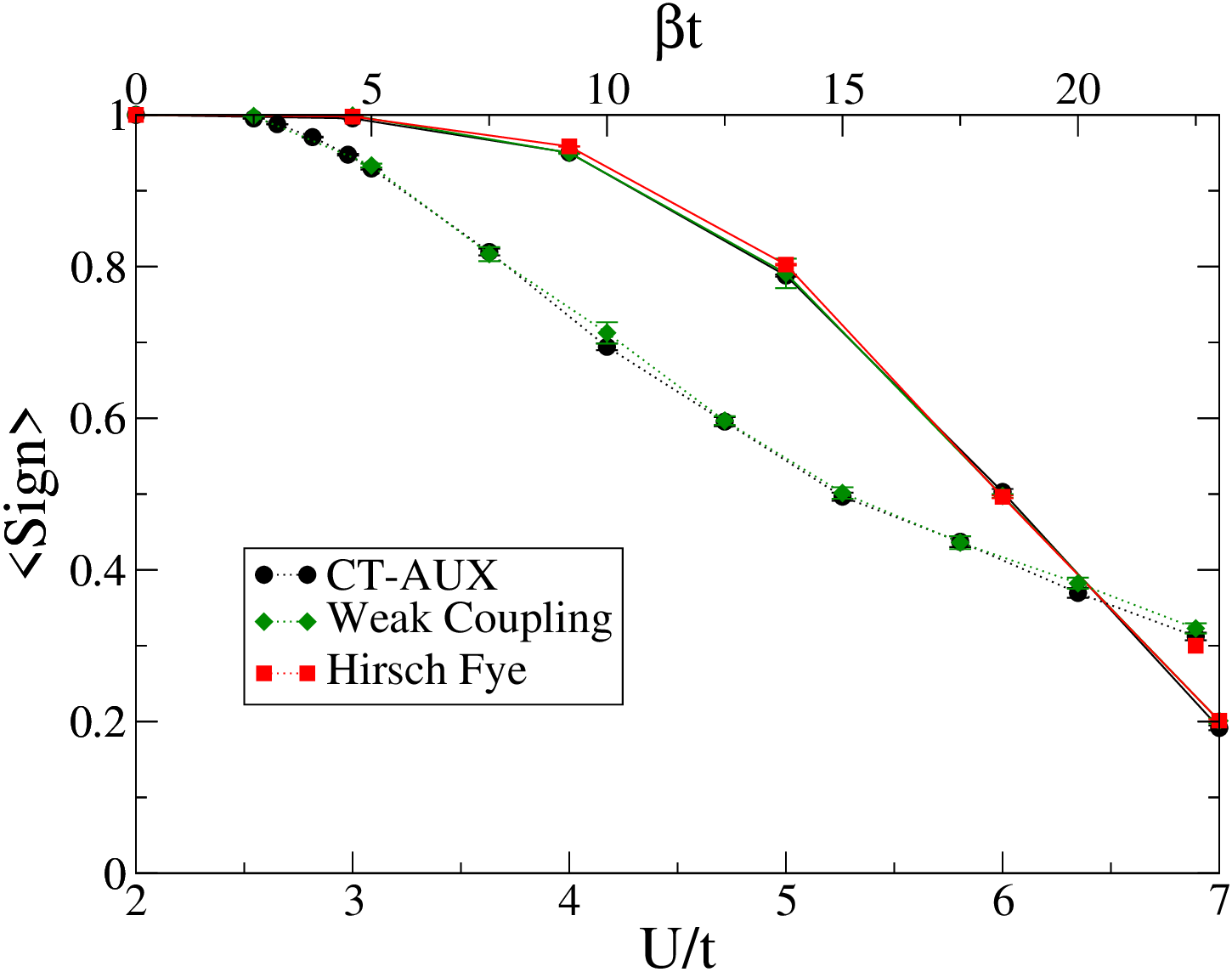}
\caption{Upper axis and dashed lines: Sign as a function of $\beta t$ for the 8-site cluster with $U = 2, \mu = -0.3757, t = 0.25.$ Lower axis and solid lines: Sign as a function of $U/t$ for the frustrated plaquette at $\beta t=10, t'/t=0.9.$}
\label{sign_combined}
\end{center}
\end{figure}

For single site impurity models, there is no sign problem since the proof 
of Ref.~\cite{Ullmo2005}
 can be extended to CT-AUX.
For cluster calculations, 
as $U$ and $\beta$ is increased, the sign becomes smaller than one. However, for the unfrustrated plaquette at temperatures down to $\beta t = 25 $ we did not observe a significant sign problem ($\langle \text{sign} \rangle \gtrsim 0.99$).
In order to produce a severe sign problem at high temperature, we frustrated our plaquette with a hopping $t'$ along the diagonal. 
For the almost triangular case $t' = 0.9 t$ the Hirsch-Fye method, weak-coupling method and our solver exhibit a sign problem that becomes increasingly severe as the interaction strength $U$ is increased or the temperature $T$ lowered.
For $U>7t$, the average sign is less than $0.2$, as seen in Fig.~\ref{sign_combined}, making it difficult to access temperatures below $T=0.1t$. 
Remarkably, the average signs in CT-AUX, \HF and the weak-coupling algorithm are almost identical.
%

Since one of the likely applications of the CT-AUX method is the solution of large DMFT clusters 
(not accessible to the hybridization expansion solver), we performed a similar study 
for an 8-site cluster, with a similar conclusion. 
The sign problem at a reference point
on the eight site Betts cluster ($U = 2,$ $t = 0.25,$ $\mu=-0.375$, $\beta=90$)\cite{ThomasPC} turned out to be the same in Hirsch-Fye as in our new algorithm (Fig. \ref{sign_combined}). 
%



\section{Conclusion}

We have presented a continuous time impurity solver based on a weak-coupling expansion of the partition function and an auxiliary field decomposition of the interaction terms. The algorithm relies on fast local updates of auxiliary Ising spin variables, whose number and position are not fixed. As a continuous time solver, our method does not suffer from the deficiencies of the Hirsch-Fye algorithm and its variants \cite{Gull07}. In particular, it does not require multiple runs for several discretizations of the imaginary time interval and subsequent extrapolations.
Moreover, it requires fewer auxiliary field variables than a standard Hirsch-Fye calculation. 
This translates into substantial gains in computational efficiency, especially
 for large clusters and at low temperature. 
 
For all regions of parameter space considered the sign problem is approximately the same in the weak coupling, Hirsch-Fye and CT-AUX algorithms.
Further investigation is however needed to determine if this is the case in all regions of parameter space and for all cluster geometries. 
We expect the new solver to be particularly useful in the simulation of large clusters, and to completely replace the Hirsch-Fye algorithm. 

\acknowledgments

The calculations have been performed on the Hreidar beowulf cluster at ETH Z{\"u}rich, using the ALPS-library,\cite{ALPS} and our cluster simulations employed a DCA self-consistency loop implemented by S.~Fuchs.  PW acknowledges support from NSF-DMR-040135.

\bibliographystyle{eplbib}
\bibliography{ctaux_paper}

\begin{thebibliography}{10}
\expandafter\ifx\csname url\endcsname\relax\def\url#1{\texttt{#1}}\fi

\bibitem{Georges96}
\Name{Georges A., Kotliar G., Krauth W. \and Rozenberg M.~J.} \REVIEW{Rev. Mod.
  Phys. }{68}{1996}{13}.

\bibitem{Georges92}
\Name{Georges A. \and Krauth W.} \REVIEW{Phys. Rev. Lett. }{69}{1992}{1240}.

\bibitem{Kotliar06}
\Name{Kotliar G., Savrasov S.~Y., Haule K. \etal} \REVIEW{Rev. Mod. Phys.
  }{78}{2006}{865}.

\bibitem{Hettler98}
\Name{Hettler M.~H., Tahvildar-Zadeh A.~N., Jarrell M. \etal} \REVIEW{Phys.
  Rev. B }{58}{1998}{R7475}.

\bibitem{Maier06}
\Name{Maier T., Jarrell M., Pruschke T. \and Hettler M.~H.} \REVIEW{Rev. Mod.
  Phys. }{77}{2005}{1027}.

\bibitem{LichtensteinKatsnelson}
\Name{{Lichtenstein} A.~I. \and {Katsnelson} M.~I.} \REVIEW{Phys. Rev. B
  }{62}{2000}{9283}.

\bibitem{Kotliar01}
\Name{Kotliar G., Savrasov S.~Y., P\'alsson G. \and Biroli G.} \REVIEW{Phys.
  Rev. Lett. }{87}{2001}{186401}.

\bibitem{Hirsch86}
\Name{Hirsch J.~E. \and Fye R.~M.} \REVIEW{Phys. Rev. Lett. }{56}{1986}{2521}.

\bibitem{Rubtsov05}
\Name{Rubtsov A.~N., Savkin V.~V. \and Lichtenstein A.~I.} \REVIEW{Phys. Rev. B
  }{72}{2005}{035122}.

\bibitem{Werner05}
\Name{Werner P., Comanac A., de' Medici L. \etal} \REVIEW{Phys. Rev. Lett.
  }{97}{2006}{076405}.

\bibitem{Werner06}
\Name{Werner P. \and Millis A.~J.} \REVIEW{Phys. Rev. B }{74}{2006}{155107}.

\bibitem{Haule07}
\Name{Haule K.} \REVIEW{Phys. Rev. B }{75}{2007}{155113}.

\bibitem{Gull07}
\Name{Gull E., Werner P., Millis A. \and Troyer M.} \REVIEW{Phys. Rev. B
  }{76}{2007}{235123}.

\bibitem{Rombouts99}
\Name{Rombouts S. M.~A., Heyde K. \and Jachowicz N.} \REVIEW{Phys. Rev. Lett.
  }{82}{1999}{4155}.

\bibitem{numrec92}
\Name{Vetterling W.~T., Flannery B.~P., Press W.~H. \and Teukolski S.~A.}
  \Book{Numerical {R}ecipes in {FORTRAN} - {T}he {A}rt of {S}cientific
  {C}omputing - {S}econd {E}dition} (University Press, Cambridge) 1992.

\bibitem{Bulla}
\Name{Bulla R., Hewson A.~C. \and Pruschke T.} \REVIEW{Journal of Physics:
  Condensed Matter }{10}{1998}{8365}.

\bibitem{Hettler00}
\Name{Hettler M.~H., Mukherjee M., Jarrell M. \and Krishnamurthy H.~R.}
  \REVIEW{Phys. Rev. B }{61}{2000}{12739}.

\bibitem{Ullmo2005}
\Name{{Yoo} J., {Chandrasekharan} S., {Kaul} R.~K. \etal} \REVIEW{Journal of
  Physics A Mathematical General }{38}{2005}{10307}.

\bibitem{ThomasPC}
\Name{Maier T.} private communication (2007).

\bibitem{ALPS}
\Name{Albuquerque A., Alet F., Corboz P. \etal} \REVIEW{Journal of Magnetism
  and Magnetic Materials }{310}{2007}{1187}.

\end{thebibliography}

\end{document}